# Stochastic Equilibrium Raman Spectroscopy (STERS)

COLBURN COBB-BRUNO[1] AND HENDRIK UTZAT*[1,2,3]

[1]*Department of Chemistry, University of California Berkeley, 200 Centennial Drive Berkeley, California 94720, USA*
[2]*Materials Science Division, Lawrence Berkeley National Lab, Berkeley, California 94720, USA*
[3]*hutzat@berkeley.edu*

**Abstract:** In this manuscript, we propose a new method for cavity- and surface-enhanced Raman spectroscopy (SERS) with improved temporal resolution in the measurement of stochastic Raman spectral fluctuations. Our approach combines Fourier spectroscopy and photon correlation to decouple the integration time from the temporal resolution. Using statistical optics Monte Carlo simulations, we establish the relationship between time resolution and Raman signal strength, revealing that typical Raman spectral fluctuations, commensurate with molecular conformational dynamics, can theoretically be resolved on micro- to millisecond timescales. The method can further extract average single-molecule dynamics from small sub-ensembles, thereby potentially mitigating challenges in achieving strictly single-molecule isolation on SERS substrates.

## 1. Introduction

Raman spectroscopy is a versatile tool widely used in chemistry, physics, and biology to determine molecular structure and phonon dispersion of materials. [1, 2] Peak *shifts* in Raman spectroscopy can also be used to measure ionic strength, temperature, and material deformation. [3, 4] Of particular interest in the chemical sciences are spectral shifts induced by intra- and intermolecular interactions, including molecular stacking, straining, or solvation. [5–9] These shifts enable the detection of molecular binding events or distinguishing between healthy and diseased states of biological tissue. [9–13] Time-resolved single-molecule Raman spectroscopy has long been proposed as a tool to access chemical and biophysical dynamics such as surface adsorption/desorption, biomolecular binding/unbinding, protein conformational changes, or stochastic molecular charging without the limitations of heterogeneous broadening. [9, 12, 14–16] This goal is limited by low Raman cross sections that require surface enhancement to achieve single-molecule sensitivity. High-enhancement plasmonic substrates can interfere with the native analyte dynamics, rendering subensemble Raman studies challenging and often inconsistent. In cases where SERS can resolve single molecules, [14, 17–26] low signal strength (Raman photons/second) compared to fluorescence typically limits the achievable time resolution of frame-by-frame SERS to the millisecond range, leaving timescales faster than the photon count rate unmeasured. [14, 17, 21, 23, 24] In contrast, ultrafast vibrational spectroscopy can reveal ensemble vibrational dynamics, for example, through two-dimensional IR spectroscopy, femtosecond stimulated Raman spectroscopy, or surface-enhanced infrared absorption spectroscopy (SEIRA). [20, 27–30] However, these methods commonly cannot discriminate between the behavior of sub-ensembles or single molecules. Importantly, both frame-by-frame SERS, ultrafast Raman, and IR methods fail to resolve nanosecond-to-millisecond dynamics, a key timescale for many molecular processes. Although recent work was able to resolve microsecond spectral shifts, this has only been achievable with the highest possible SERS enhancement factors from extreme field confinement in plasmonic gap cavities. [21, 31, 32] Such plasmonic gap cavities are unsuitable for the investigation of unperturbed temporal fluctuations in the molecular Raman shift of larger analytes, for example, single biomacromolecules or polymers. Thus, new methods to reveal fast single-molecule dynamics from low-enhancement SERS structures are needed.

Here, we present a theoretical framework for a new SERS method addressing two current challenges: (1) the limited time resolution imposed by low SERS signals and (2) difficulty discriminating between single- and few-molecule Raman spectral dynamics. This paper introduces our new method, Stochastic Equilibrium Raman Spectroscopy (STERS), and provides a theoretical and numerical analysis of its analytical figures of merit. Experimentally, STERS analyzes temporal correlations in Raman-scattered photon energies via a scanning Michelson interferometer. In ergodic systems, temporal correlation functions quantify average changes between two time points, and in spectroscopy, the spectral correlation function measures how much, on average, an optical spectrum shifts in frequency over a given period. Therefore, as long as Raman spectral fluctuations are ergodic over the course of the experiment, the spectral correlation function $p(\tau, \zeta)$ can be used to quantify the timescale and amplitude of spectral dynamics, where $\tau$ is the lag time between photons, $\zeta$ is the energy difference between two scattered photons, and $\langle ... \rangle$ denotes a time average:

$$p(\tau, \zeta) = \left\langle \int_{-\infty}^{\infty} s(t, \omega) s(t + \tau, \omega + \zeta) d\omega \right\rangle. \tag{1}$$

Then, spectral diffusion timescales are quantified via the evolution of the spectral correlation with respect to $\tau$, as simulated in Figure 1 (simulation details are discussed in the results section). Figure 1 b illustrates how the dynamic timescale can be extracted by measuring the width of the spectral correlation as a function of $\tau$. By using photon correlations, STERS gains the general advantages of high spectral and temporal resolution/dynamic range, figures of merit that are fundamentally constrained in grating-based Raman spectroscopy.

This concept was previously introduced as Photon-Correlation Fourier Spectroscopy (PCFS) to study spectral diffusion of the electronic transitions in single emitters. [33] In short, time-stamped photon detection is performed with two photon counting detectors at the output arm of a scanning Michelson interferometer (Figure SI 1). The $g^2(\tau)_\delta$ auto- and cross-correlation functions are determined at different interferometer path length differences ($\delta$) across the emitter's coherence length. Then, the Fourier Transform of $g^2(\tau)_\delta$ from path length difference to energy difference ($\zeta$) gives the spectral correlation $p(\zeta, \tau)$. While PCFS has been used to study different systems at cryogenic and room temperature, [34–36] we propose that extending this approach to single-molecule Raman spectroscopy is particularly powerful, as it enables quantification of vibrational spectral diffusion as an indirect measure of local molecular dynamics.

To support the viability of our new method, a key question is whether vibrational spectral diffusion can be resolved under typical experimental conditions. We address this through numerical simulations to determine the expected temporal resolution for parameterized spectral diffusion models and low photon count rates. We then evaluate their relevance to practical experimental constraints and propose STERS as a promising approach for measuring multi-timescale Raman spectral diffusion.

## 2. Simulation Results

We parameterize stochastic Raman spectral fluctuations using two phenomenological models and perform statistical optics Monte Carlo simulations with experimentally informed parameters for the Raman lineshape and signal strength. These models describe the discrete stochastic switching of a Raman line between different center frequencies, characterized by a time constant $\tau_c$. [37, 38] These simulations assume shot-noise-limited measurements across varying photon count rates.

In the two-state model (Figure 2 b, inset), the Lorentzian center frequency alternates between two spectrally distinct states, representing scenarios such as switching between two

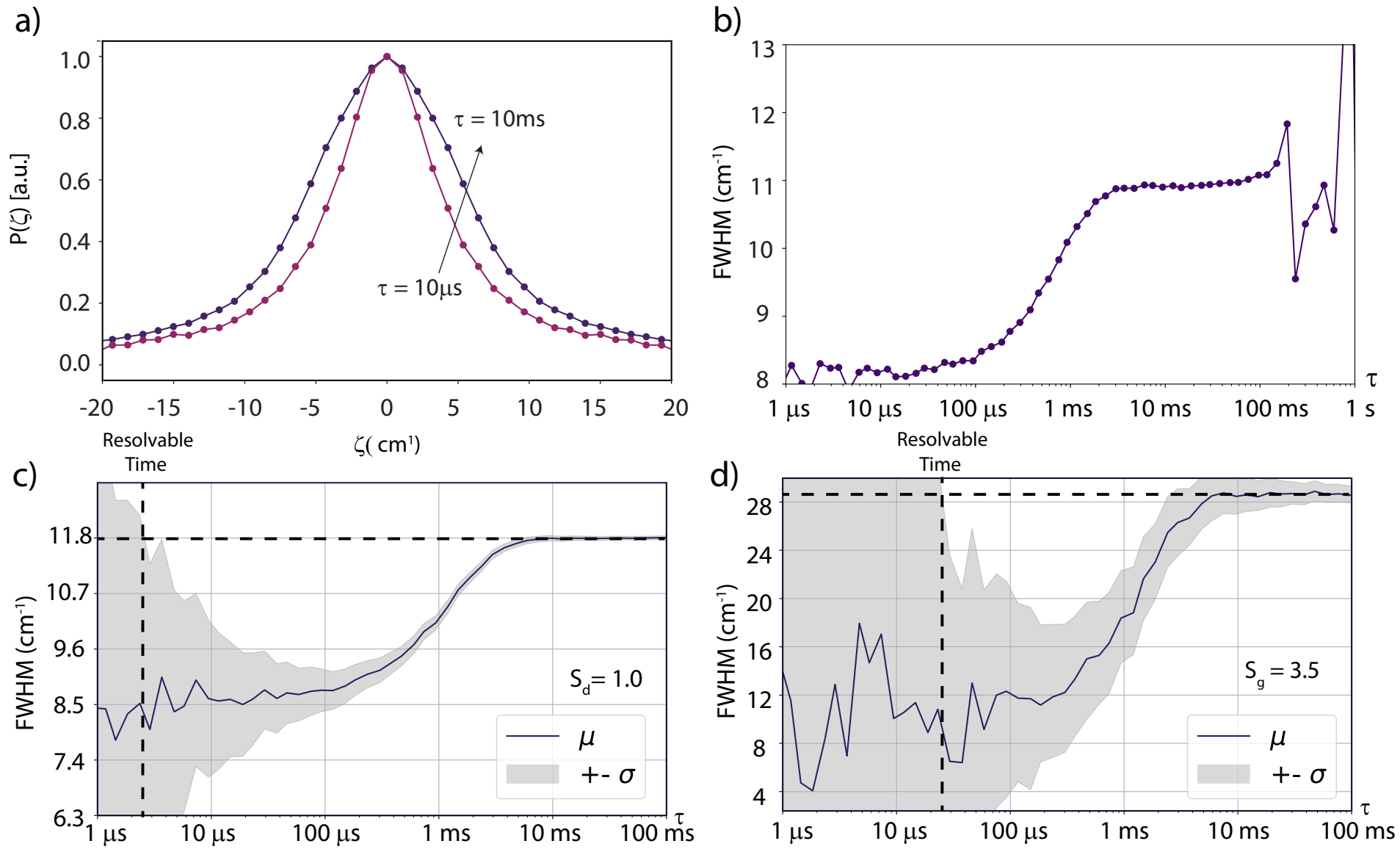

Fig. 1. Simulated plots introduce the spectral correlation and the definition of resolvability. (a) shows the spectral correlation at select delay times ($\tau$). (b) shows the evolution of the FWHM with $\tau$ illustrating a spectral fluctuation at 1 ms. c) and d) show the evolution of the FWHM with $\tau$ for an experimental time of about 5 hours, including standard deviations obtained via 100 consecutive simulations. c) uses $S_d = 1$ and 1000 photons per second and d) uses $S_g = 3.5$ with 100 photons per second.

discrete conformations. In this model, line broadening evolves from the intrinsic lineshape

$$s(\omega) = \frac{\frac{1}{\Gamma}}{\frac{1}{\Gamma^2} + (\omega - \omega_{1,2})^2}, \tag{2}$$

to the diffused lineshape

$$s(\omega) = \frac{\frac{1}{\Gamma}}{\frac{1}{\Gamma^2} + (\omega - \omega_1)^2} + \frac{\frac{1}{\Gamma}}{\frac{1}{\Gamma^2} + (\omega - \omega_2)^2}. \tag{3}$$

In the Gaussian model (Figure 2 a, inset), discrete spectral jumps occur within a Gaussian probability distribution over frequency, mimicking a continuous distribution of possible states. This model results in a broadened Voigt lineshape described by the integral

$$s(\omega) = \int_{-\infty}^{\infty} e^{-\frac{1}{2}\left(\frac{\omega' - \mu}{\sigma}\right)^2} \frac{\frac{1}{\Gamma}}{\frac{1}{\Gamma^2} + (\omega - \omega')^2} d\omega', \tag{4}$$

with the intrinsic lineshape

$$s(\omega) = \frac{\frac{1}{\Gamma}}{\frac{1}{\Gamma^2} + (\omega - \omega')^2}. \tag{5}$$

Both models are parameterized in their Lorentzian linewidth $\Gamma$. For the Gaussian model, the width of the spectral jumping probability distribution is defined by $\sigma$. For the two-state model, the spectral separation between the two states is given by $\Delta\nu = |\omega_1 - \omega_2|$, functionally replacing $\sigma$. As in other spectral shift measurements, the resolvability of STERS depends on the ratio between the intrinsic and broadened linewidths. To quantify dynamic linewidth broadening, we introduce the parameters $S_d \equiv \frac{\Delta\nu}{\Gamma}$ for the two-state model and $S_g \equiv \frac{\sigma}{\Gamma}$ for the Gaussian model as measures of the dynamic linewidth broadening. These parameters effectively measure how much the intrinsic Raman lineshape is broadened due to spectral diffusion. The stochastic model for spectral diffusion can be found in [39].

In the following sections, we simulate the spectral correlation for select cases of $S_{d,g}$. We then discuss the results in the context of estimated $S_d$ and $S_g$ derived from published experimental data.

## 2.1. Quantification of Raman Spectral Diffusion Using STERS

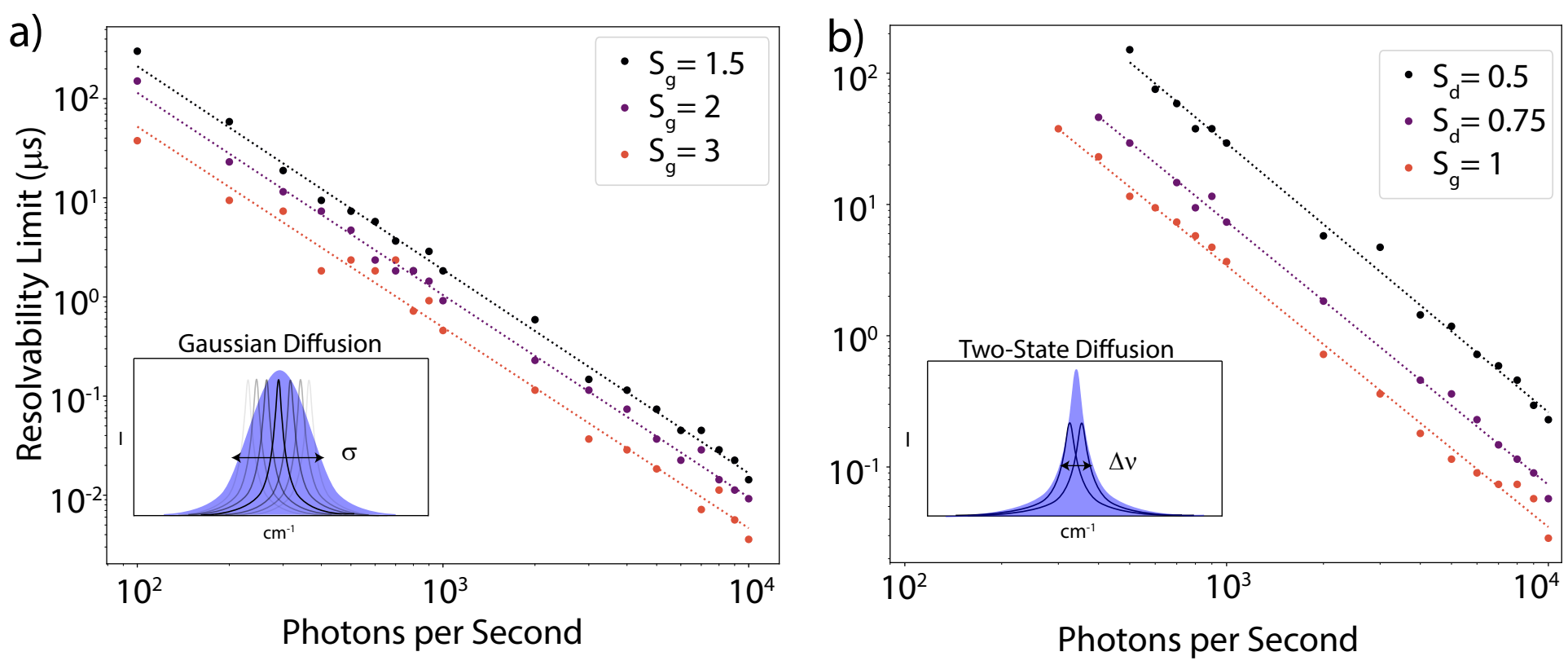

Fig. 2. Simulated resolvability as a function of different values of $S_d, g$. Each data point represents the average result from 100 independent simulations, each corresponding to five hours of experimental time. The insets at the bottom left illustrate the diffusion model used in each case, providing a visual representation of the underlying dynamics.

We first explore the ability of STERS to resolve small magnitude spectral fluctuations ($0.5 < S_{d,g} < 3$) of a single, isolated Raman line undergoing spectral diffusion. Our main objective is to identify the achievable time resolution at low photon count rates for different values of $S_{d,g}$. In brief, Poisson-distributed photon arrival times are simulated, with photon frequencies drawn from statistical distributions and mapped onto virtual detectors based on frequency and interferometer position. [39] STERS analysis is then performed on this virtual photon-counting data. We chose a center Raman mode frequency of 1106 $cm^{-1}$, characteristic of an aromatic ring breathing mode. This choice of the Raman Stokes wavelength affects the delay stage motion parameters in our simulations, but it does not impact the final spectral correlation, which only depends on the linewidth and spectral diffusion, as the spectral correlation is always centered at $\zeta = 0$. The intrinsic linewidth $\Gamma$ used in the simulations is 4 $cm^{-1}$, an estimate from [18, 40]. The details of these simulations are discussed in the Methods section.

Representative simulation results are shown in Figure 1 c and d. The gray region around the mean represents the standard deviation of the full width at half maximum (FWHM) of $p(\zeta, \tau)$, obtained from 100 independent simulations. Spectral diffusion may still be resolved as long as the observed spectral correlation broadening exceeds the uncertainty of the measurement. For our purposes, we define the achievable time resolution for a given $S_d, g$ as the value of $\tau$ at which the FWHM uncertainty starts to exceed the observed broadening. This is illustrated in Figure 1 c,d where the resolvability would be $\sim 4\mu$s and $\sim 15\mu$s and is indicated by vertical dashed lines.

The achievable resolvability a shot-noise limited STERS measurement depends on the total number of photon pairs collected for a given $\tau$. Our simulations assess the achievable resolution for different $S_{g,d}$ with reasonably long total experiment times (five hours) and variable rates of Raman signal photons per second per molecule (100 to 10,000). The chosen $S_g$ values for the Gaussian model are 1.5, 2, and 3 (Figure 2), and $S_d$ values for the two-state model are 0.5, 0.75, and 1 (Figure 2 b). Values of $S_d > 1$ are excluded, as the spectral correlation adopts a triplet lineshape, requiring different fitting of the spectral correlation and revised definition of the shown FWHM. This is justified by our focus on low-amplitude spectral diffusion. The resolvability of a fluctuating Raman mode that follows a Gaussian diffusion model ranges from $\sim$100 $\mu$s with 100 Raman photons per second per molecule to $\sim$10 ns with 10,000 photons per second per molecule in our chosen range of $S_g$. Comparing with $S_d$ in Figure 2 b reveals a tenfold lower time resolution for similar photon counts for the two-state model, thereby defining a range for the expected resolvability for different diffusion models.

### 2.2. Resolving Multi-Timescale Dynamics

Furthermore, STERS can resolve temporally distinct spectral fluctuations in the same spectral line, for example of the form $s(\omega, t) = s(\omega)_0 + \delta s(\omega(t))_1 + \delta s(\omega(t))_2$, where the two fluctuation terms $\delta s(\omega(t))_{1,2}$ can possess different amplitudes and characteristic diffusion times. We illustrate this capability through simulations of two independent spectral diffusion processes with $S_d = 0.25$ and 0.5 and characteristic diffusion times $\tau_c$ of $10\mu s$ and $1ms$ shown in Figure 3. We propose that similar parameters could arise from a scatterer traversing a potential energy surface with four wells, each corresponding to a different mode frequency associated with four forward and backward transition rates. As in fluorescence correlation spectroscopy (FCS), only the faster of each forward and backward transition rate pair appears in the spectral correlation because rapid broadening of $p(\zeta, \tau)$ obscures slower broadening of equal amplitude. [41] As a result, the FWHM of $p(\zeta, \tau)$ evolves with two distinct fluctuation times, $\tau_{c1}$ and $\tau_{c2}$, shown in Figure 3 b.

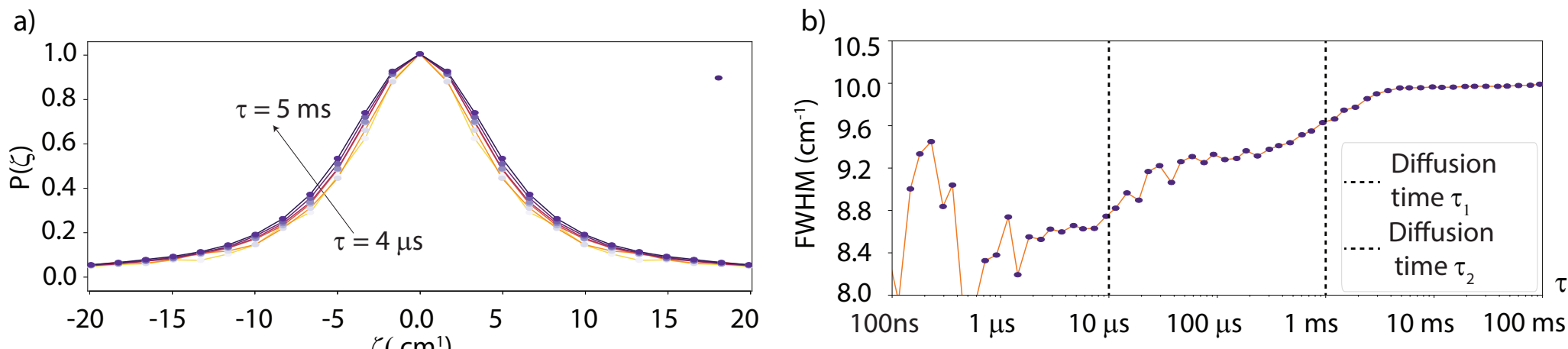

Fig. 3. Simulated plots illustrating two fluctuation timescales. (a) shows the spectral correlation at select delay times ($\tau$). (b) presents the evolution of the FWHM over time. The simulation was performed with a photon count rate of 1000 photons per second, an intrinsic Lorentzian linewidth of 4 $cm^{-1}$, and two discrete spectral jumps of $S_d = 0.25$ and 0.5, occurring with diffusion times of $\tau_c = 10\mu s$ and 1 ms.

### 2.3. Obtaining Average Single-Molecule Raman Dynamics from Sub-Ensembles

Isolating, verifying, and measuring SERS from a single molecule/analyte remains a challenge using conventional techniques. [42] Like FCS, STERS can measure average single-molecule dynamics from a sub-ensemble, thereby eliminating the effect of inhomogeneous broadening. This avoids both the difficulty of strictly single-molecule observation and may allow for more consistent sample preparation, for example, by using larger mode volume nanocavities hosting sub-ensembles of molecules. As the correlation of an ensemble of independent spectral diffusers separates into autocorrelation and cross-correlation components, the total spectral correlation can be written in terms of ensemble and single-molecule components:

$$p(\zeta, \tau) = p_{\text{single-molecule}}(\zeta, \tau) + p_{\text{ensemble}}(\zeta), \tag{6}$$

where $p_{\text{ensemble}}(\zeta)$ is $\tau$-invariant. It can thus be subtracted from the total to isolate the $\tau$-dependent $p_{\text{single-molecule}}(\zeta, \tau)$. We provide further derivations and explanations in the SI. The signal-to-background ratio (SBR) decreases with the number N of molecules under observation. With $p_{\text{ensemble}}(\zeta)$ as the background and $p_{\text{single-molecule}}(\tau, \zeta)$ as the signal, the SBR can be expressed as:

$$SBR = \frac{p_{\text{single-molecule}}(\zeta, \tau)}{p_{\text{ensemble}}(\zeta)} = \frac{1}{N-1}, \tag{7}$$

which is undefined at N = 1 due to the absence of any background terms. This relationship also applies to solution phase FCS. [43] The experimental observable for a small ensemble is shown in Figure 4, comparing the spectral correlation for a single-molecule (N=1) and a small ensemble (N=4) with independently diffusing emitters ($S_g$ = 2). A comparison between Figure 4 a and Figure 4 c reveals that for N=1 and N=4, the spectral correlation width increases by $\simeq 8cm^{-1}$ and $\simeq 2.4cm^{-1}$ which corresponds to the SBR decrease of $\frac{1}{3}$. The smaller relative increase observed for larger ensembles reflects the growing contribution of the $\tau$-invariant $p_{\text{ensemble}}(\zeta)$ component of the spectral correlation.

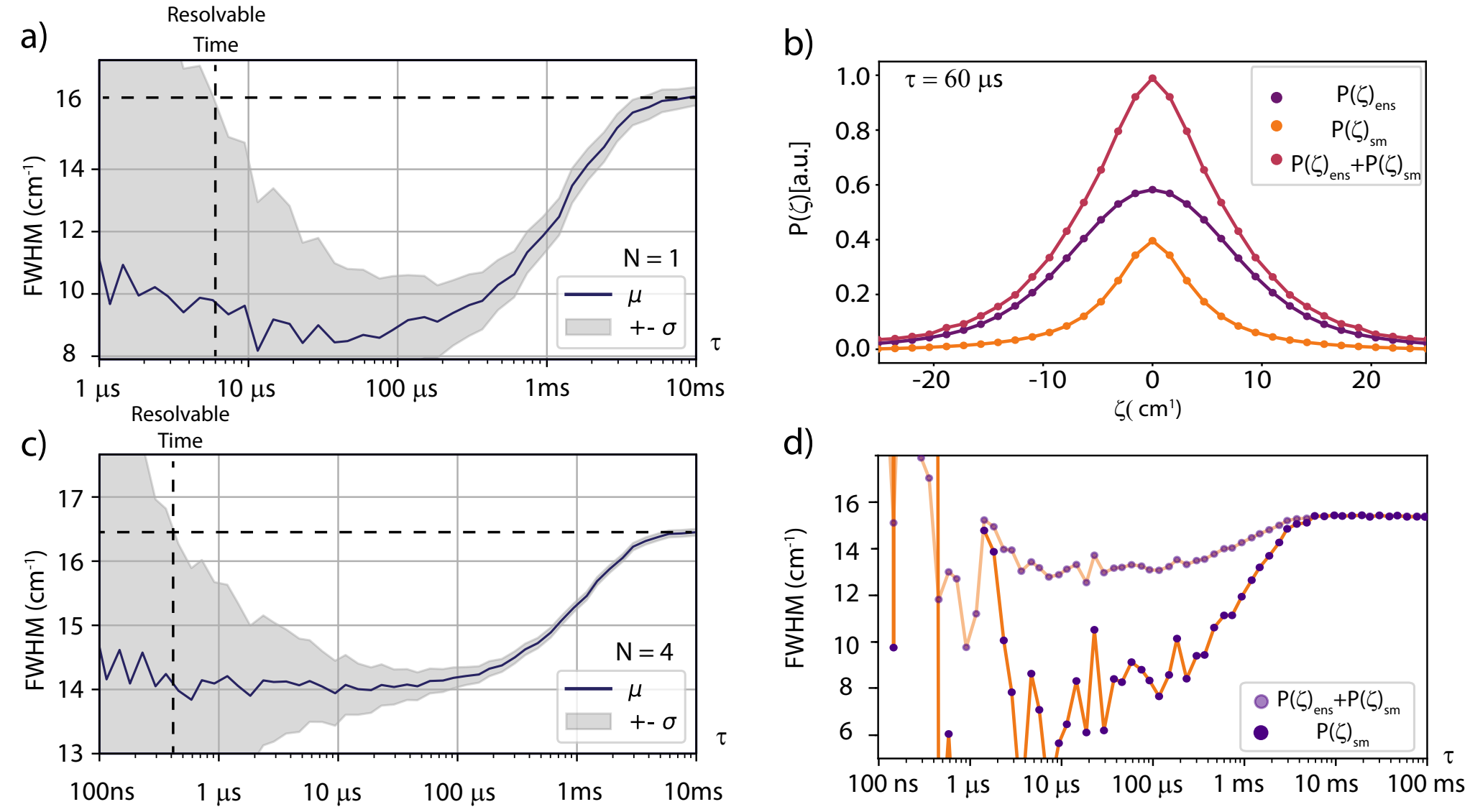

Fig. 4. Evolution of the spectral correlation over delay time ($\tau$) for ensemble size N = 1 (a) and N = 4 (c), with a correlation time $\tau_c$ = 1ms. a) and c) were simulated with 10,000 photons per second for a total experimental time of 30 minutes. Means and error bars represent the standard deviation across 100 independent simulations. b) shows the spectral correlation for an ensemble size N = 4 fitted to a diffused Voigt profile and a Lorentzian, representing the ensemble and the single-molecule spectral correlation given the relationship in Eq. (7). d) shows the evolution of the FWHM for the ensemble and single-molecule spectral correlation components from b). b) and d) were simulated with a total experimental time of 5 hours. All simulations were performed with a fixed spectral shift parameter of $S_g$ = 2.

By decomposing the total spectral correlation into ensemble and single-molecule components, the single-molecule contribution $p_{\text{single-molecule}}(\zeta, \tau)$ can be extracted from the sum spectral correlation. Figure 4 b shows this decomposition into $p_{\text{ensemble}}(\zeta)$ with a Voigt lineshape profile and $p_{\text{single-molecule}}(\tau, \zeta)$ with a Lorentzian shape. Figure 4 d compares the $\tau$-dependent FWHM of the single-molecule component extracted from the total spectral correlation. Notably, although the amplitude of $p_{\text{single-molecule}}(\tau, \zeta)$ decreases with increasing N, the shot noise on both $p_{\text{single-molecule}}(\tau, \zeta)$ and $p_{\text{ensemble}}(\zeta)$ decreases due to a higher total photon count rate. Therefore, $p_{\text{single-molecule}}(\tau, \zeta)$ can, in principle, be obtained for high N. [43] While we can define a relationship for the SBR of the spectral correlation, a relationship for the SNR of the spectral correlation is not only dependent on the count rate of the signal, but also dependent on the number of stage positions used. This is because more stage positions allow for the white noise from the correlation functions to be spread over a higher frequency range.

### *2.4. Multi-Dimensional Cross-Correlation STERS*

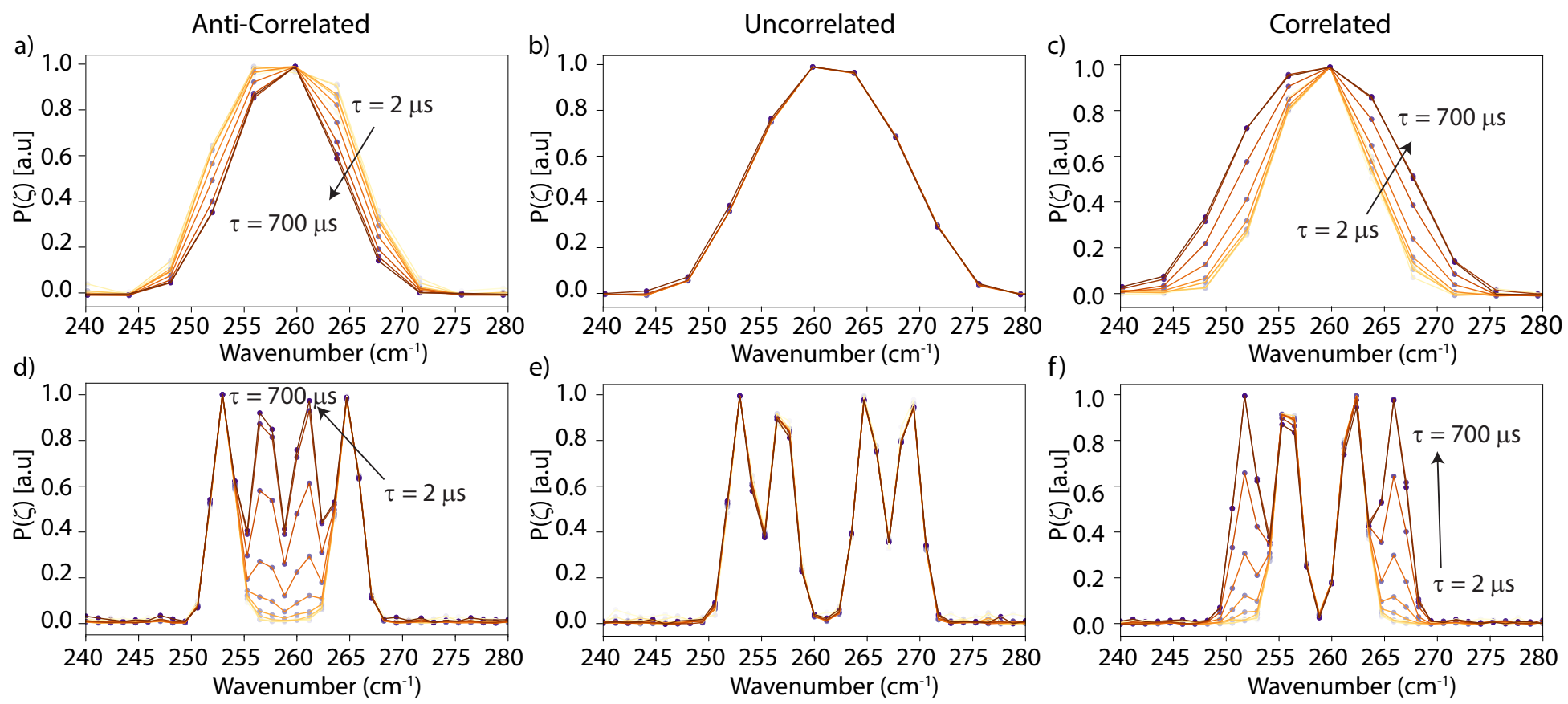

Fig. 5. Illustration of the multi-dimensional STERS observable. Columns (left to right) represent anti-correlated, uncorrelated, and correlated frequency fluctuations. Two Raman modes at 847 and 1106 $cm^{-1}$ were considered. a)-c) were simulated with an intrinsic linewidth of 4 $cm^{-1}$ with switching frequency magnitudes of $4cm^{-1}$ and $5cm^{-1}$ ($S_d$ = 2.67, 3.33) and a characteristic fluctuation time of $\tau_c = 100\mu s$. c)-e) were simulated with a linewidth of 1.5 $cm^{-1}$ with the same switching parameters ($S_d$ = 1, 1.25).

The preceding discussion focuses on diffusion within a single Raman spectral line. However, multi-dimensional correlation analysis across different Raman lines naturally provides more information about system fluctuations. Distinctly correlated, anti-correlated, or uncorrelated frequency fluctuations have been observed on timescales of seconds and picoseconds. [44, 45] STERS enables resolution of similar cross-correlations on intermediate timescales. Figure 5 illustrates how correlated, uncorrelated, and anti-correlated Raman spectral fluctuations between two modes manifest in the STERS spectral correlation. As an example, we consider two phenomenological modes at 847 and 1106 $cm^{-1}$, corresponding to the C=O stretching and ring-breathing modes of tyrosine. [46] In general, the qualitative features of $p(\zeta, \tau)$ remain similar for any two modes with different absolute center energies.

Figure 5 d-f illustrates the general form of the multi-dimensional spectral correlation. For two modes that are anti-correlated, the spectral correlation evolves from a wide doublet into a quartet (Figure 5 d), whereas for two correlated modes, it evolves from a narrow doublet into a quartet (Figure 5 f). Two uncorrelated modes show no change, as seen in Figure 5 e. At smaller linewidth/shift ratios, the spectral correlation appears as a single peak that broadens under correlated fluctuations and narrows under anti-correlated ones (Figure 5 a,c). As before, uncorrelated modes show no change. This behavior arises because, for anti-correlated modes, the energy difference is maximal before $\tau_c$ and decreases for $\tau > \tau_c$, narrowing the spectral correlation. For correlated modes, the trend reverses, leading to broadening of the spectral correlation.

## 3. Discussion

STERS builds on a previous method, Photon Correlation Fourier Spectroscopy (PCFS), and represents its first application to vibrational spectroscopy. [33] Based on our simulations, we identify several distinct advantages of STERS and contextualize them by comparison with existing techniques and proposed mechanisms of vibrational spectral diffusion.

A key advantage of STERS lies in its use of interferometry and photon correlation. This approach circumvents the trade-off between spectral resolution and signal throughput that occurs with grating-based spectrometers. Furthermore, because STERS achieves temporal resolution through energy-time correlation analysis (at the cost of absolute photon energy), it fills the current time gap in single-molecule vibrational spectroscopy from nanoseconds to milliseconds where pump-probe and frame-by-frame methods are limited. On these timescales, fluctuations in the local electric field strength, adsorption geometry, biomolecular conformational structure, and molecular interactions contribute to dynamic Raman linewidth broadening. [3, 4, 9–13] While single-molecule studies that definitively measure these mechanisms are limited, a range for expected values of $S_{d,g}$ may still be considered.

Upper limits for the expected $S_g$ may be estimated from inhomogeneous broadening. At low temperatures, variations in the Raman mode center energies between different single molecules exceed the homogeneous linewidth by a factor of two to five. [14, 18] At room temperature, inhomogeneous broadening decreases by between one and two and a half times the homogeneous width. [14, 18, 40] For example, Kang et al. observe ~7.5 $cm^{-1}$ frequency variations in TERS modes of adsorbed molecules compared to linewidths of ~6 $cm^{-1}$. [47] Assuming similar dynamic broadening, this corresponds to $2 < S_g < 5$ at low temperatures and $1 < S_g < 2.5$ at room temperature. These values should be seen as upper limits, because the spectral diffusion amplitude is expected to be lower than the inhomogeneous broadening.

Comparable values for $S_g$ can also be predicted based on the relationship between molecular forces and frequency shifts in anharmonic vibrational modes. Considering the model laid out by Hutchinson et al.. Force fluctuations of ≃ 25-100 pN can result in $\Delta\nu$ shifts of $2 - 5cm^{-1}$ for aromatic ring breathing, nitrile stretching, and CO stretching modes, which corresponds to $0.4 < S_g < 2.5$ for typical single-molecule Raman linewidths of $2 - 5cm^{-1}$. [14, 18, 40] These $S_g$ values are based on the relationship $\Delta\nu \simeq -F\frac{3\nu_e g}{2f^2}$, where $\nu_e$ is the unperturbed frequency and $f$ and $g$ are the bond's harmonic and anharmonic force constants, respectively. [48] It has also been reported that fluctuating forces on the order of piconewtons can emerge from protein binding, folding, and conformational switching, [49–57] with rare instances of nanonewton forces also reported. [58]

The above values for $S_g$ are also consistent with studies on force-transduced Raman shifts in different environments. Kho et al. report force-induced SERS shifts of ~ 2–4 $cm^{-1}$ from molecular crowding and Mostafa et al. observe the 1269 $cm^{-1}$ mode of cytochrome C fluctuating between two values separated by 12 $cm^{-1}$ on the timescale of seconds due to a still unknown mechanism. [9, 15]

As seen in Figure 2, spectral diffusion with $1 < S_d < 3$ can be resolved on the micro- to millisecond timescale, depending on the count rate and total integration time, reaching far into the temporal blind spot of current methods. SERS photon count rates, although not always directly reported, typically range from 100 to 1,000,000 per second per molecule, the highest of which are reported for 'pico-cavities' formed by atomic protrusions. [21,31,32] The achievable time resolution of STERS is thus highly system-dependent, ranging from hundreds of nanoseconds to milliseconds. The experimental feasibility of STERS, however, depends on prolonged experimental times, which is particularly challenging in the case of the highest enhancement plasmonic substrates with bright single-molecule SERS signals that often suffer from sample instability. STERS can mitigate this by enabling lower excitation powers or by using more stable, lower-enhancement substrates that reduce local heating at the expense of tighter field confinement. In this vein, STERS may enable the measurement of SERS dynamics using dielectric nanophotonics for enhancement, which has shown promise due to low heat conversion and minimal background. [59,60].

STERS further distinguishes itself through its ability to cross-correlate different Raman spectral lines to form multidimensional spectral fluctuation maps. This multi-dimensional approach provides insights similar to correlation analyses previously demonstrated in TERS, but extends across a wider range of timescales. [14] Park et al. proposed that surface adsorption dynamics can be understood through correlation analysis, where molecular orientation on metallic surfaces anti-correlates modes with different symmetry and correlates those with the same symmetry. We propose testing multi-dimensional STERS to analyze macromolecular conformational dynamics and intermolecular forces, revealing solvent exposure and force transduction between different parts of the molecule. This approach could complement 2D NMR, which lacks single-molecule sensitivity, and 2D IR spectroscopy, which has limited time resolution. [61–67]

Finally, we highlight that STERS eliminates the requirement to isolate strictly one molecule in the SERS hotspot. Similar to FCS, STERS can extract average single-molecule dynamics even for small sub-ensembles. [68] This aspect helps to overcome the common issue in SERS of reliable single-molecule placement and verification using the bi-analyte method, which is prone to problems from differences in dual-analyte adsorption affinities or molecular grouping. [42,69–71] While STERS does not resolve individual spectral trajectories, it captures ensemble-averaged single-molecule dynamics, removes static inhomogeneous broadening, and offers a path to addressing the broader reproducibility challenges of high-enhancement plasmonic substrates.

## 4. Conclusion

In conclusion, we propose a new Raman method to analyze vibrational spectral diffusion using the spectral auto- and cross-correlations between different Raman lines. The advantages of STERS may help to address long-standing challenges in SERS and offer a more accessible and reliable means of quantifying vibrational spectral diffusion, which will be of interest for both fundamental and applied studies of molecular dynamics. We envision STERS to bridge the gap between the frame-by-frame SERS approach and ultrafast vibrational spectroscopy, specifically the difficult-to-reach range of nanoseconds to milliseconds, important for the characterization of biomolecular dynamics, [72–74] single-molecule SERS [14,17,21,28], and catalysis. [23,24]

## Supporting Info

## 5. Methods: Numerical Simulations

A stream of photons, simulated with Poisson time statistics, is assigned a detector arrival time and energy based on its arrival time relative to frequency information about the emitter. The emitter frequency is assigned for all of the experimental time with a probability of switching to another value at a time $\tau_c$. Then, a probability of being detected at detector 1 or 2 is assigned based on the path length difference and photon energy. Once the stream of photons has all of the information about energy, detector, and arrival time, the detectors are correlated using an algorithm laid out in [75]. With a $g^{(2)}_{\delta}(\tau)$ function for each stage position $\delta$, a $\tau$ is chosen to create the $g^{(2)}_{\tau}(\delta)$ interferogram, which is Fourier transformed to get the spectral correlation function for all $\tau$. It is important to note that the true values for the FWHM may be slightly less or slightly higher than the simulated value. This is due to the number of stage positions chosen. More stage positions can be chosen to increase the spectral resolution, but this must be balanced with considerations of experimental time. this can lead to a slight under-sampling of the interferogram, which can also alter the true values of the linewidths.
We draw each photon from a Lorentzian probability distribution, $\text{FWHM}_1$, with some center frequency, and then after the fluctuation time, we choose a different center frequency with a probability distribution defined by a Gaussian of linewidth, $\text{FWHM}_2$, where the representative parameter is $S_g$ as defined above.

## 6. Derivation of SBR

To derive the Signal to Background (SBR) ratio of STERS, the spectral correlation can first be written as:

$$p(\tau,\zeta) = \left\langle \int_{-\infty}^{\infty} s(t,\omega)s(t+\tau,\omega+\zeta)d\omega \right\rangle \equiv \left\langle s(t,\omega) * s(t+\tau,\omega+\zeta) \right\rangle. \tag{8}$$

where condensed notation for the convolution is used. In a sub-ensemble of independent spectrally diffusing emitters, the overall spectrum $(s(t,\omega))$ is a linear combination of multiple scatterers which allows the spectral correlation to be written like this:

$$\left\langle s(t,\omega) * s(t+\tau,\omega+\zeta) \right\rangle = \left\langle \left( \sum_{i=1}^{n} s_i(t,\omega) \right) * \left( \sum_{j=1}^{n} s_j(t+\tau,\omega+\zeta) \right) \right\rangle. \tag{9}$$

The product under the integral can be multiplied out giving separated spectral auto-correlations and cross-correlations:

$$\left\langle s(t,\omega) * s(t+\tau,\omega+\zeta) \right\rangle = \left\langle \sum_{i=j} s_i(t,\omega) * s_j(t+\tau,\omega+\zeta) \right\rangle + \left\langle \sum_{i\neq j} s_i(t,\omega) * s_j(t,+\tau,\omega+\zeta) \right\rangle. \tag{10}$$

This is equivalent to

$$p(\tau,\zeta) = p_{\text{single-molecule}}(\tau,\zeta) + p_{\text{ensemble}}(\tau,\zeta). \tag{11}$$

The single-molecule spectral correlation $p_{\text{single-molecule}}(\tau,\zeta)$ and the ensemble spectral correlation $p_{\text{ensemble}}(\tau,\zeta)$ add linearly to the total spectral correlation. From this point, it can be seen that there are n terms where i=j, therefore:

$$\text{signal} = N. \tag{12}$$

Now there also are $N^2$ total terms, and since there are only the two cases where i=j and i≠j, there must be $N^2 - N$ terms such that i≠j. These are the background terms.

$$\text{background} = N^2 - N \tag{13}$$

This shows that the amplitude of $p_{\text{single-molecule}}(\tau,\zeta)$ relative to $p_{\text{ensemble}}(\tau,\zeta)$ or the SBR decreases with increasing number N of simultaneously measured molecules with the relation

$$\frac{\text{signal}}{\text{background}} = \frac{p_{\text{single-molecule}}(\tau,\zeta)}{p_{\text{ensemble}}(\tau,\zeta)} = \frac{1}{N-1}. \tag{14}$$

## 7. Experimental Setup

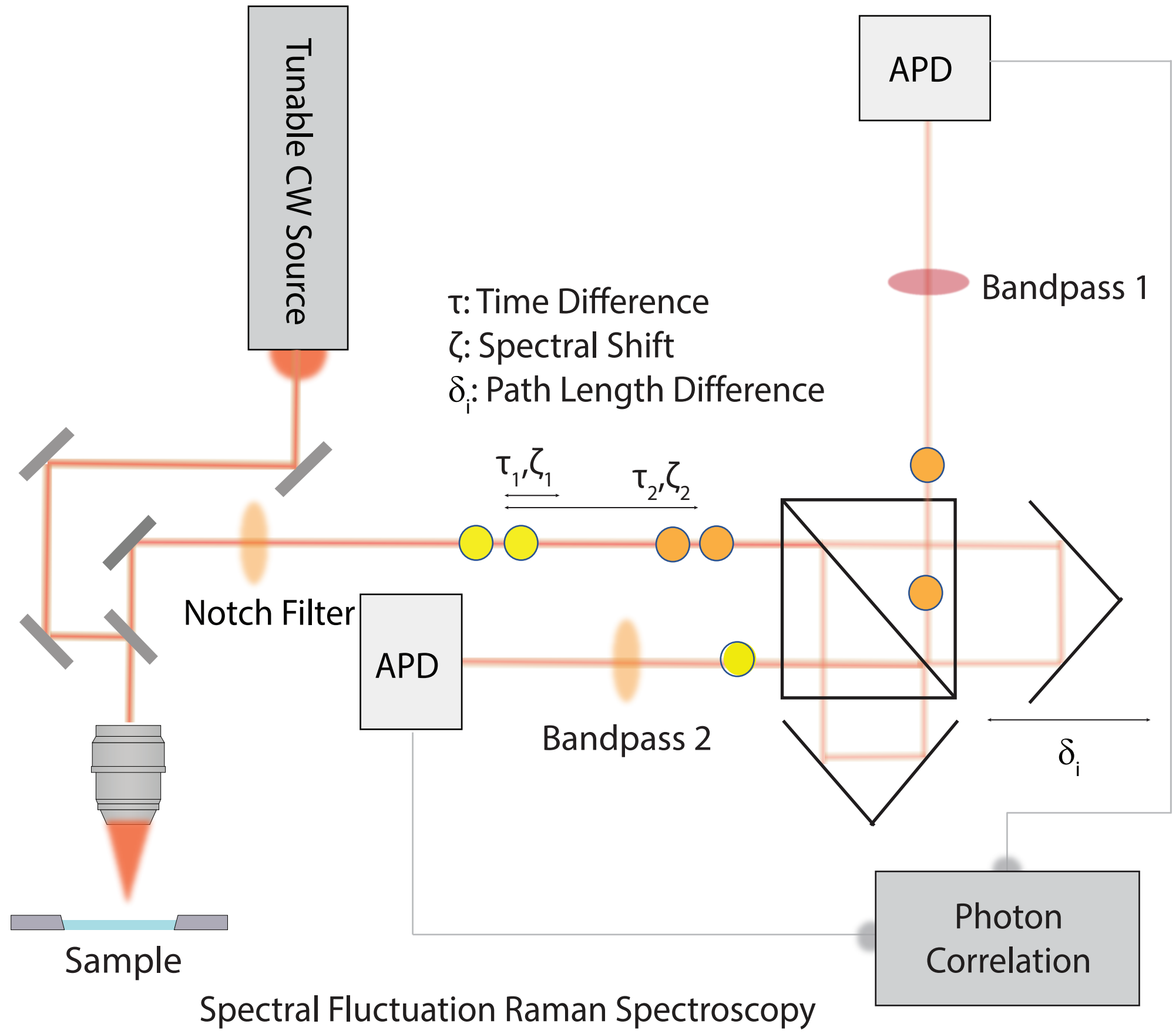

Fig. 6. Experimental schematic that includes a confocal microscope with its output coupled to a Michelson interferometer. The output of the interferometer is directed to single-photon avalanche diodes with bandpass filters to select a Raman mode.

## References

1. D. C. Smith and C. Carabatos-Nédelec, "Raman spectroscopy applied to crystals: phenomena and principles, concepts and conventions," PRACTICAL SPECTROSCOPY SERIES **28**, 349–422 (2001).
2. J. R. Ferraro, *Introductory raman spectroscopy* (Elsevier, 2003).
3. S. Prawer and R. J. Nemanich, "Raman spectroscopy of diamond and doped diamond," Philos. Trans. Royal Soc. London. Ser. A: Math. Phys. Eng. Sci. **362**, 2537–2565 (2004).
4. V. Zani, D. Pedron, R. Pilot, and R. Signorini, "Contactless temperature sensing at the microscale based on titanium dioxide raman thermometry," Biosensors **11**, 102 (2021).
5. W. R. Welch, J. Kubelka, and T. A. Keiderling, "Infrared, vibrational circular dichroism, and raman spectral simulations for $\beta$-sheet structures with various isotopic labels, interstrand, and stacking arrangements using density functional theory," The J. Phys. Chem. B **117**, 10343–10358 (2013).

6. W. Nuansing, A. Rebollo, J. M. Mercero, *et al.*, “Vibrational spectroscopy of self-assembling aromatic peptide derivates,” J. Raman Spectrosc. **43**, 1397–1406 (2012).
7. L. Sparks, W. R. Scheidt, and J. Shelnutt, “Effects of. pi.-. pi. interactions on molecular structure and resonance raman spectra of crystalline copper (ii) octaethylporphyrin,” Inorg. Chem. **31**, 2191–2196 (1992).
8. T. Yamaguchi, Y. Kimura, and N. Hirota, “Solvent and solvent density effects on the spectral shifts and the bandwidths of the absorption and the resonance raman spectra of phenol blue,” The J. Phys. Chem. A **102**, 2252–2252 (1998).
9. K. W. Kho, U. Dinish, A. Kumar, and M. Olivo, “Frequency shifts in sers for biosensing,” ACS nano **6**, 4892–4902 (2012).
10. J. T. Motz, S. J. Gandhi, O. R. Scepanovic, *et al.*, “Real-time raman system for in vivo disease diagnosis,” J. biomedical optics **10**, 031113–031113 (2005).
11. D. Kurouski, R. P. Van Duyne, and I. K. Lednev, “Exploring the structure and formation mechanism of amyloid fibrils by raman spectroscopy: a review,” Analyst **140**, 4967–4980 (2015).
12. W. Zhu, J. A. Hutchison, M. Dong, and M. Li, “Frequency shift surface-enhanced raman spectroscopy sensing: An ultrasensitive multiplex assay for biomarkers in human health,” ACS sensors **6**, 1704–1716 (2021).
13. B. Tang, J. Wang, J. A. Hutchison, *et al.*, “Ultrasensitive, multiplex raman frequency shift immunoassay of liver cancer biomarkers in physiological media,” ACS nano **10**, 871–879 (2016).
14. K.-D. Park, E. A. Muller, V. Kravtsov, *et al.*, “Variable-temperature tip-enhanced raman spectroscopy of single-molecule fluctuations and dynamics,” Nano letters **16**, 479–487 (2016).
15. A. Mostafa, Y. Kanehira, K. Tapio, and I. Bald, “From bulk to single molecules: Surface-enhanced raman scattering of cytochrome c using plasmonic dna origami nanoantennas,” Nano Lett. (2024).
16. A. M. Rao, P. Eklund, S. Bandow, *et al.*, “Evidence for charge transfer in doped carbon nanotube bundles from raman scattering,” Nature **388**, 257–259 (1997).
17. H.-H. Shin, G. J. Yeon, H.-K. Choi, *et al.*, “Frequency-domain proof of the existence of atomic-scale sers hot-spots,” Nano letters **18**, 262–271 (2018).
18. C. Artur, E. C. Le Ru, and P. G. Etchegoin, “Temperature dependence of the homogeneous broadening of resonant raman peaks measured by single-molecule surface-enhanced raman spectroscopy,” The J. Phys. Chem. Lett. **2**, 3002–3005 (2011).
19. J. M. Klingsporn, N. Jiang, E. A. Pozzi, *et al.*, “Intramolecular insight into adsorbate–substrate interactions via low-temperature, ultrahigh-vacuum tip-enhanced raman spectroscopy,” J. Am. Chem. Soc. **136**, 3881–3887 (2014).
20. L. A. Jakob, W. M. Deacon, R. Arul, *et al.*, “Accelerated molecular vibrational decay and suppressed electronic nonlinearities in plasmonic cavities through coherent raman scattering,” Phys. Rev. B **109**, 195404 (2024).
21. M. M. Schmidt, E. A. Farley, M. A. Engevik, *et al.*, “High-speed spectral characterization of single-molecule sers fluctuations,” ACS nano **17**, 6675–6686 (2023).
22. N. Chiang, X. Chen, G. Goubert, *et al.*, “Conformational contrast of surface-mediated molecular switches yields ångstrom-scale spatial resolution in ultrahigh vacuum tip-enhanced raman spectroscopy,” Nano letters **16**, 7774–7778 (2016).
23. C. M. Zoltowski, D. N. Shoup, and Z. D. Schultz, “Investigation of sers frequency fluctuations relevant to sensing and catalysis,” The J. Phys. Chem. C **126**, 14547–14557 (2022).
24. H.-K. Choi, W.-H. Park, C.-G. Park, *et al.*, “Metal-catalyzed chemical reaction of single molecules directly probed by vibrational spectroscopy,” J. Am. Chem. Soc. **138**, 4673–4684 (2016).
25. P. G. Etchegoin and E. C. Le Ru, “Resolving single molecules in surface-enhanced raman scattering within the inhomogeneous broadening of raman peaks,” Anal. chemistry **82**, 2888–2892 (2010).
26. S. Yampolsky, D. A. Fishman, S. Dey, *et al.*, “Seeing a single molecule vibrate through time-resolved coherent anti-stokes raman scattering,” Nat. Photonics **8**, 650–656 (2014).
27. A. Tokmakoff and M. Fayer, “Homogeneous vibrational dynamics and inhomogeneous broadening in glass-forming liquids: Infrared photon echo experiments from room temperature to 10 k,” The J. chemical physics **103**, 2810–2826 (1995).
28. H. Arnolds and M. Bonn, “Ultrafast surface vibrational dynamics,” Surf. Sci. Reports **65**, 45–66 (2010).
29. A. E. Bragg, W. Yu, J. Zhou, and T. Magnanelli, “Ultrafast raman spectroscopy as a probe of local structure and dynamics in photoexcited conjugated materials,” The journal physical chemistry letters **7**, 3990–4000 (2016).
30. P. Kukura, D. W. McCamant, and R. A. Mathies, “Femtosecond stimulated raman spectroscopy,” Annu. Rev. Phys. Chem. **58**, 461–488 (2007).
31. N. C. Lindquist, C. D. L. de Albuquerque, R. G. Sobral-Filho, *et al.*, “High-speed imaging of surface-enhanced raman scattering fluctuations from individual nanoparticles,” Nat. nanotechnology **14**, 981–987 (2019).
32. M. Kamp, B. de Nijs, N. Kongsuwan, *et al.*, “Cascaded nanooptics to probe microsecond atomic-scale phenomena,” Proc. National Acad. Sci. **117**, 14819–14826 (2020).
33. X. Brokmann, M. Bawendi, L. Coolen, and J.-P. Hermier, “Photon-correlation fourier spectroscopy,” Opt. express **14**, 6333–6341 (2006).
34. H. Utzat, W. Sun, A. E. Kaplan, *et al.*, “Coherent single-photon emission from colloidal lead halide perovskite quantum dots,” Science **363**, 1068–1072 (2019).
35. B. Spokoyny, H. Utzat, H. Moon, *et al.*, “Effect of spectral diffusion on the coherence properties of a single quantum emitter in hexagonal boron nitride,” The journal physical chemistry letters **11**, 1330–1335 (2020).
36. J. Cui *et al.*, “Deconstructing the room-temperature emission spectra of nanocrystals using photon-correlation fourier

spectroscopy," Ph.D. thesis, Massachusetts Institute of Technology (2014).
37. M. Peters, T. Zhao, S. George, *et al.*, "Energy landscape of conformational changes for a single unmodified protein," npj Biosensing **1**, 14 (2024).
38. W. Ye, M. Gotz, S. Celiksoy, *et al.*, "Conformational dynamics of a single protein monitored for 24 h at video rate," Nano letters **18**, 6633–6637 (2018).
39. H. Utzat and M. G. Bawendi, "Lifetime-resolved photon-correlation fourier spectroscopy," Opt. Express **29**, 14293–14303 (2021).
40. S. George, A. Harris, M. Berg, and C. Harris, "Picosecond studies of the temperature dependence of homogeneous and inhomogeneous vibrational linewidth broadening in liquid acetonitrile," (1983).
41. F. A. Nia, *Probing Molecular Kinetics Using Higher-Order Fluorescence Correlation Spectroscopy* (Colorado State University, 2019).
42. E. Blackie, E. Le Ru, M. Meyer, *et al.*, "Bi-analyte sers with isotopically edited dyes," Phys. Chem. Chem. Phys. **10**, 4147–4153 (2008).
43. S. Saffarian and E. L. Elson, "Statistical analysis of fluorescence correlation spectroscopy: the standard deviation and bias," Biophys. journal **84**, 2030–2042 (2003).
44. M. D. Sonntag, D. Chulhai, T. Seideman, *et al.*, "The origin of relative intensity fluctuations in single-molecule tip-enhanced raman spectroscopy," J. Am. Chem. Soc. **135**, 17187–17192 (2013).
45. N. Demirdöven, M. Khalil, O. Golonzka, and A. Tokmakoff, "Correlation effects in the two-dimensional vibrational spectroscopy of coupled vibrations," The J. Phys. Chem. A **105**, 8025–8030 (2001).
46. S. A. Bhat and S. Ahmad, "Ftir, ft-raman and uv–vis spectral studies of d-tyrosine molecule," J. Mol. Struct. **1105**, 169–177 (2016).
47. M. Kang, H. Kim, E. Oleiki, *et al.*, "Conformational heterogeneity of molecules physisorbed on a gold surface at room temperature," Nat. Commun. **13**, 4133 (2022).
48. E. J. Hutchinson and D. Ben-Amotz, "Molecular force measurement in liquids and solids using vibrational spectroscopy," The J. Phys. Chem. B **102**, 3354–3362 (1998).
49. C. J. Bustamante, Y. R. Chemla, S. Liu, and M. D. Wang, "Optical tweezers in single-molecule biophysics," Nat. Rev. Methods Primers **1**, 25 (2021).
50. E.-L. Florin, V. T. Moy, and H. E. Gaub, "Adhesion forces between individual ligand-receptor pairs," Science **264**, 415–417 (1994).
51. M. S. Kellermayer, S. B. Smith, H. L. Granzier, and C. Bustamante, "Folding-unfolding transitions in single titin molecules characterized with laser tweezers," Science **276**, 1112–1116 (1997).
52. M. Rief, M. Gautel, F. Oesterhelt, *et al.*, "Reversible unfolding of individual titin immunoglobulin domains by afm," science **276**, 1109–1112 (1997).
53. J.-Y. Shao, H. P. Ting-Beall, and R. M. Hochmuth, "Static and dynamic lengths of neutrophil microvilli," Proc. National Acad. Sci. **95**, 6797–6802 (1998).
54. T. R. Strick, J.-F. Allemand, D. Bensimon, *et al.*, "The elasticity of a single supercoiled dna molecule," Science **271**, 1835–1837 (1996).
55. T. Strick, J.-F. Allemand, D. Bensimon, and V. Croquette, "Stress-induced structural transitions in dna and proteins," Annu. review biophysics biomolecular structure **29**, 523–543 (2000).
56. L. Tskhovrebova, J. Trinick, J. Sleep, and R. Simmons, "Elasticity and unfolding of single molecules of the giant muscle protein titin," Nature **387**, 308–312 (1997).
57. M. Akamatsu, R. Vasan, D. Serwas, *et al.*, "Principles of self-organization and load adaptation by the actin cytoskeleton during clathrin-mediated endocytosis," Elife **9**, e49840 (2020).
58. M. Jasnin, J. Hervy, S. Balor, *et al.*, "Elasticity of podosome actin networks produces nanonewton protrusive forces," Nat. Commun. **13**, 3842 (2022).
59. J. Cambiasso, M. Konig, E. Cortes, *et al.*, "Surface-enhanced spectroscopies of a molecular monolayer in an all-dielectric nanoantenna," Acs Photonics **5**, 1546–1557 (2018).
60. M. Caldarola, P. Albella, E. Cortés, *et al.*, "Non-plasmonic nanoantennas for surface enhanced spectroscopies with ultra-low heat conversion," Nat. communications **6**, 7915 (2015).
61. L. E. Kay, D. A. Torchia, and A. Bax, "Backbone dynamics of proteins as studied by nitrogen-15 inverse detected heteronuclear nmr spectroscopy: application to staphylococcal nuclease," Biochemistry **28**, 8972–8979 (1989).
62. A. G. Palmer III, M. Rance, and P. E. Wright, "Intramolecular motions of a zinc finger dna-binding domain from xfin characterized by proton-detected natural abundance carbon-13 heteronuclear nmr spectroscopy," J. Am. Chem. Soc. **113**, 4371–4380 (1991).
63. N. A. Farrow, R. Muhandiram, A. U. Singer, *et al.*, "Backbone dynamics of a free and a phosphopeptide-complexed src homology 2 domain studied by 15n nmr relaxation," Biochemistry **33**, 5984–6003 (1994).
64. M. T. Zanni, "Two-dimensional infrared spectroscopy measures the structural dynamics of a self-assembled film only one molecule thick," Proc. National Acad. Sci. **113**, 4890–4891 (2016).
65. J. Lim, K.-K. Lee, C. Liang, *et al.*, "Two-dimensional infrared spectroscopy and molecular dynamics simulation studies of nonaqueous lithium ion battery electrolytes," The J. Phys. Chem. B **123**, 6651–6663 (2019).
66. D. C. Urbanek, D. Y. Vorobyev, A. L. Serrano, *et al.*, "The two-dimensional vibrational echo of a nitrile probe of the villin hp35 protein," The journal physical chemistry letters **1**, 3311–3315 (2010).
67. Z. Ganim, H. S. Chung, A. W. Smith, *et al.*, "Amide i two-dimensional infrared spectroscopy of proteins," Accounts

chemical research **41**, 432–441 (2008).
68. S. T. Hess, S. Huang, A. A. Heikal, and W. W. Webb, "Biological and chemical applications of fluorescence correlation spectroscopy: a review," Biochemistry **41**, 697–705 (2002).
69. P. G. Etchegoin, E. C. Le Ru, and A. Fainstein, "Bi-analyte single molecule sers technique with simultaneous spatial resolution," Phys. Chem. Chem. Phys. **13**, 4500–4506 (2011).
70. J. A. Dieringer, R. B. Lettan, K. A. Scheidt, and R. P. Van Duyne, "A frequency domain existence proof of single-molecule surface-enhanced raman spectroscopy," J. Am. Chem. Soc. **129**, 16249–16256 (2007).
71. E. C. Le Ru, M. Meyer, and P. G. Etchegoin, "Proof of single-molecule sensitivity in surface enhanced raman scattering (sers) by means of a two-analyte technique," The journal physical chemistry B **110**, 1944–1948 (2006).
72. A. Abyzov, M. Blackledge, and M. Zweckstetter, "Conformational dynamics of intrinsically disordered proteins regulate biomolecular condensate chemistry," Chem. Rev. **122**, 6719–6748 (2022).
73. P. Campitelli, T. Modi, S. Kumar, and S. B. Ozkan, "The role of conformational dynamics and allostery in modulating protein evolution," Annu. review biophysics **49**, 267–288 (2020).
74. J. Guo and H.-X. Zhou, "Protein allostery and conformational dynamics," Chem. reviews **116**, 6503–6515 (2016).
75. T. A. Laurence, S. Fore, and T. Huser, "Fast, flexible algorithm for calculating photon correlations," Opt. letters **31**, 829–831 (2006).